\documentclass{sig-alternate}

\usepackage{url}

\usepackage{breakurl}
\usepackage[breaklinks]{hyperref}

\begin{document}

\title{Anomaly Detection in the Bitcoin System - A Network Perspective}

\numberofauthors{2} 
\author{
\alignauthor
Thai T. Pham\thanks{Graduate School of Business, Stanford University. Email: \texttt{thaipham@stanford.edu}. We would like to thank Jure Leskovec for helpful suggestions and comments.}
\alignauthor
Steven Lee\thanks{Computer Science Department, Stanford University. Email: \texttt{slee2010@stanford.edu}}
}

\maketitle
\begin{abstract}
The problem of anomaly detection has been studied for a long time, and many Network Analysis techniques have been proposed as solutions. Although some results appear to be quite promising, no method is clearly to be superior to the rest. In this paper, we particularly consider anomaly detection in the Bitcoin transaction network. Our goal is to detect which users and transactions are the most suspicious; in this case, anomalous behavior is a proxy for suspicious behavior. To this end, we use the laws of power degree $\&$ densification and local outlier factor (LOF) method (which is proceeded by $k$-means clustering method) on two graphs generated by the Bitcoin transaction network: one graph has users as nodes, and the other has transactions as nodes. 

We remark that the methods used here can be applied to any type of setting with an inherent graph structure, including, but not limited to, computer networks, telecommunications networks, auction networks, security networks, social networks, Web networks, or any financial networks. We use the Bitcoin transaction network in this paper due to the availability, size, and attractiveness of the data set. 
\end{abstract}


\section{Introduction}
Network structures have appeared for a long time, and along with them are those who behave abnormally within the system. We refer to these people or their illegal activities as anomalies. With respect to financial transactional networks, anomalies can include those who execute fraudulent transactions. In these networks, a common goal is to detect those anomalies to prevent future illegal actions. People really care about detecting the anomalies due to increasing theft rates, both specifically in the Bitcoin network and in other financial networks.

Bitcoin is a special type of transaction system; more information about it can be found in \url{http://bitcoin.org/en/}. We seek to use anomaly detection as a proxy for suspicious users/activities detection. Similarly, anomaly detection is used to detect fradulent purchases in credit card systems or network congestion in computer networks. However, in this anonymous network, nodes (i.e. users, transactions) are unlabeled and there is no confirmation as to whether or not a given node is actually conducting illicit activities.

In this project, we focus particularly on the problem of detecting anomalies in the Bitcoin transaction network, which is certainly related to the study of fraud detection in all types of financial transaction systems in which a rich literature is available (Pham and Lee \cite{PL2016}, Farren et al. \cite{FPA2016}). Since this problem can be generalized to those in other network settings, which may or may not involve financial transactions, we are examining the more general problem of anomaly detection in networks. 

The rest of the paper is organized as follows. Section $2$ presents related work. Section $3$ describes our methods, including data collection and parsing, feature extraction, and mathematical descriptions of the network analysis techniques. Section $4$ discusses evaluation methods. Section $5$ presents the results we obtain by running these techniques on the network-type data set we generated. Section $6$ evaluates our methods and results. Section $7$ mentions future works. Section $8$ concludes our study.   

\section{Related Work}
There are many research studies that concern anomaly detection, and these studies use a variety of techniques including machine learning and network analysis techniques. For the purpose of this paper, we will review related network analysis results only.

Leskovec et al. \cite{LKF2007} describe the properties of graph evolution over time, which are different from what many previous papers would predict. One key idea is the densification power law which states that in log - log scale, most graphs densify with the number of edges being a linear function of the number of nodes. Leskovec et al. then discuss Forest Fire Model, which reproduces many key properties of real networks including densification power law, heavy tailed in/out - degrees, etc. The comment given out is that the graphs are abnormal if these laws and characteristics fail to satisfy. This is the motivation for our first analysis that uses densification power law and power law degree distribution on the Bitcoin network. We will also use the power law degree distribution for other features other than in-degree and out-degree. If we detect any structures induced by users' activities that deviate significantly from these laws, we may conclude that there is something abnormal in the network. We will use this as the preliminary anomaly detection method to see if anomalous structures actually exist in the network.

Smith et al. \cite{SBEPS2002} make use of clustering techniques to detect anomalies. The main idea is that these methods should be able to group normal users/activities together and separate from abnormal ones. Smith et al. use k-means clustering, self-organizing maps, and the expected maximization algorithm to develop methods for the detection process. Motivated by this, we think we can have a good use of k-means clustering method on the Bitcoin dataset. However, we will not use k-means as a real method to detect anomalies because clustering in its deep sense is for grouping purposes. Instead, we will use it as a baseline model. Since we expect outliers (i.e. anomalies) to stay far away from the centroids found by $k$-means, $k$-means can be used to assess our true method. In the same sense, $k$-means is helpful for visualization purposes. Most importantly, without $k$-means to find the centroids, we cannot calculate LOF indices in the next part, which defines our notion of anomalies; the connection between $k$-means clustering method and LOF method will be discussed in more details in the Methods section. 

Breunig et al. \cite{BKNS2000} propose the Local Outlier Factor (LOF) method to detect outliers in a dataset. This method relies on the concept of local density, with locality defined by $k$ nearest neighbors and density estimated by distances. Basically, they compare the local density of a point (a node) to that of its neighbors to identify regions of similar density and points that have substantially higher densities than their neighbors; these points are then labeled as outliers. We find this method suitable for our study because not only outliers can be understood as anomalies, but also we do not need labelled data to feed in the calculations. Thus, we will use this method as our main methodology to detect anomalies in the Bitcoin network. We then use k-means clustering results discussed above to verify our findings.       

\section{Methods}
In this section, we will first data collection and parsing. Then, we describe feature extraction, and finally we provide mathematical explanations for the social network analysis techniques we will use. 

\subsection{Data Collection and Parsing}
We use the Bitcoin transaction data set obtained from Stanford Network Analysis Project. All Bitcoin transactions are documented in a public ledger and are in the currency unit called the Bitcoin (BTC). The data set contains all Bitcoin transactions beginning from the network's creation until April 7th, 2013. For each transaction, there can be multiple sender and receiver addresses. Furthermore, multiple addresses can belong to a single user. Finally, users are also anonymous in that there are no names or personal information associated with a given user. 

The data set is quite large: there are $6,336,769$ users with $37,450,461$ transactions. We parse the data in two ways. The first way, which we will call the \emph{user graph}, is quite intuitive: users (where each user owns a list of addresses) are nodes and transactions between users are edges. The second way, which we will call the \emph{transaction graph}, models transactions as nodes and Bitcoin flow between transactions as edges.

In our analysis, we will use both graph types to investigate the Bitcoin network. The user graph will help us detect suspicious users, while the transaction graph will help us detect suspicious transactions. Using these two graph representations, we can not only find out both abnormal users and abnormal activities, but also check if our methods are consistent in the sense that suspicious transactions should belong to suspicious users.   

\subsection{Feature Extraction}
In order to use $k$-means as a baseline and calculate the local outlier factor (LOF), for each node in the graph we extract a set of features. Note that for the two afore-mentioned network representations of the data, the set of features are not the same; even if they share the same feature names, those names may mean very different things.

\subsubsection{User Graph}
For the user graph, we choose the set of features from the followings:

\begin{itemize}
  \item In-degree: Number of transactions received by a given user.
  \item Out-degree: Number of transactions sent by a given user.
  \item Unique in-degree: Number of unique users a given user has received transactions from.
  \item Unique out-degree: Number of unique users a given user has sent transactions to.
  \item Average in-transaction: Average number of bitcoins received per incoming transaction.
  \item Average out-transaction: Average number of bitcoins sent per outgoing transaction.
  \item Average time interval between in-transactions.
  \item Average time interval between out-transactions.
  \item Number of public keys owned by a given user.
  \item Balance: Net number of bitcoins retained by user.
  \item Clustering coefficient: measure of connectivity amongst neighbors of a given user.
  \item Creation date: timestamp of first transaction associated with a given user.
  \item Active duration: time difference between first and most recent transactions associated with a given user.
  
  \item \ldots
\end{itemize}

\subsubsection{Transaction Graph}
For the transaction graph, we choose the set of features from the followings:

\begin{itemize}
  \item In-degree: Number of transactions (i.e. nodes) that tak money from a given transaction (i.e. a given node). 
  \item Out-degree: Number of transactions that a given transaction takes money from.
  \item Unique in-degree: Number of unique transactions that take money from a given transaction.
  \item Unique out-degree: Number of unique transactions that a given transaction takes money from. 
  \item Average in-transaction: Average number of bitcoins on each incoming edge to a given transaction.
  \item Average out-transaction: Average number of bitcoins on each outgoing edge from a given transaction.
  \item Average time interval between in-transactions.
  \item Average time interval between out-transactions.
  \item Number of users a given transaction is associated with.
  \item Balance: Net number of bitcoins for a given transaction considering all in- and out-going edges from that transaction. 
  \item Clustering coefficient: measure of connectivity amongst neighbors of a given transaction.
  \item Creation date: timestamp of first edge associated with a given transaction. 
  \item Active duration: time difference between first and most recent edges associated with a given transaction.
  
  \item \ldots
\end{itemize}

\subsection{Network Analysis Techniques}
We will use three methods to discover and detect anomalies in the bitcoin network. These methods should be also suited for anomaly detection in any network setting systems. 

\subsubsection{Power Degree $\&$ Densification Laws}
Generally, the densification power law states that
\begin{equation*} E(t) \propto N(t)^\alpha, \end{equation*}
where $N(t)$ and $E(t)$ are numbers of nodes and edges at time $t$, and $\alpha$ is some positive constant. The power law degree states that in many real world network,
\begin{equation*} P(k) \propto k^{-\gamma}, \end{equation*}
where $P(k)$ is the fraction of nodes that have degree $k$ and $\gamma$ is some positive constant. Hence in a normal network, $\log(P(k))$ is a linear function of $\log(k)$. We expect this relation to hold when nodes' degrees are replaced by their in-degrees, their out-degrees, and in the bitcoin system that we consider, their balances, average transaction sizes, and total amounts of transactions as well.    

\subsubsection{$k$-means Clustering}
The purpose of the $k$-means clustering method is to partition $m$ points (i.e. $m$ nodes in the graph) into $k$ groups of similar characteristics. The optimal choice of $k$ for the Bitcoin data will be determined in the Results section.

For this method to work, we first need to represent each node as a multi-dimensional vector in the Euclidean space; each dimension of a node is a feature that we choose from the list described in part $2.2$. For runtime purposes, we did not use the entire list of features per node; instead we select a subset of features for each node. For the user-as-node graph, we use the following six features.

\begin{itemize}
    \item In-degree
    \item Out-degree
    \item Mean incoming transaction value
    \item Mean outgoing transaction value
    \item Mean time interval
    \item Clustering coefficient
\end{itemize}

For the transaction-as-node network, we use three features:

\begin{itemize}
    \item In-degree
    \item Out-degree
    \item Total value of the transaction
\end{itemize}

This procedure produces a set of $m$ points (i.e. nodes) $(x_1, ..., x_m)$ for which $x_i \in \mathbb{R}^n$ for each $i = 1, ..., m$. Here $n = 6$ or $3$. We seek to partition these $m$ points into $k$ clusters $S = (S_1, ..., S_k)$ to solve 
\begin{equation*} \min_S \sum_{i = 1}^k \sum_{x \in S_i} ||x - \mu_i||^2, \end{equation*}
where $\mu_i$ is the mean of the points in $S_i$ for each $i = 1, ..., k$. We use the $k$-means clustering algorithm as a heuristic method to solve this problem. The algorithm in details can be found in Lloyd \cite{L1982}.

Our final note in this part is that to account for the power law nature of the feature values, we use the normalized log of feature values.

\subsubsection{Local Outlier Factor}
We denote by $d(A, B)$ the distance between two points $A$ and $B$. Furthermore, we denote by $k$-distance(A) the distance of the point $A$ to its $k$-th nearest neighbor. We also define the set of $k$-th nearest neighbors of a point $A$:
\begin{equation*} N_k(A) = \{B \; | \; d(A, B) = k\text{-distance(A)} \}. \end{equation*}

Then, we define \emph{reachability distance}:
\begin{equation*}                                     \text{reachability-distance}_k(A, B) = \max\{k\text{-distance(B)}, d(A, B)\}. \end{equation*}

In other words, the \emph{reachability distance} indexed $k$ of a point $A$ from a point $B$ is the true distance between the two points, but this distance must be at least the $k$-distance of $B$. Note that the reachability distance is not symmetric. Then, we can define the most important definition in this part, the \emph{local reachability density} of a point $A$, to be 
\begin{equation*} 
    LRD(A) := \left( \frac{\sum_{B \in N_k(A)} \text{reachability-distance}_k(A, B)}{|N_k(A)|} \right)^{-1}. 
\end{equation*}

Intuitively, it is the inverse of the average reachability distance of the point $A$ from its neighbors. Then, we use these quantities to calculate $LOF_k(A)$, which is defined by the average local reachability density of $A$'s neighbors divided by the $A$'s own local reachability density, i.e.
\begin{equation*} 
    LOF_k(A) = \frac{\sum_{B \in N_k(A)} LRD(B)}{|N_k(A)|} \backslash LRD(A).
\end{equation*}

So if $LOF_k(A) \approx 1$ for some point $A$, then we imply that $A$ is comparable to its neighbors and therefore $A$ is not an outlier; if $LOF_k(A) << 1$, then we imply a denser region and we conclude that $A$ might be an inlier; and if $LOF_k(A) >> 1$, we indicate $A$ as an outlier. 

A final remark in this part is about what we meant earlier about using $k$-means clustering to calculate the LOF indices. Technically, we do not need the results obtained by $k$-means clustering to determine the LOF indices; instead, we need $k$-nearest-neighbors. When implementing the LOF method, however, we use $k$-means clustering results to narrow down the list of potential $k$-nearest-neighbors for a given node. Basically, using k-means, once we obtain the list of nodes for each cluster, we can (somewhat reasonably) search for a given node's $k$-nearest-neighbors within the cluster it is grouped in - this saves computational time by a factor of $\sim 49$ for computing the LOF indices.

\section{Evaluation Methods}
With unlabeled data, evaluating our methods is a major challenge. Due to the nature of the network-type data set we have, we propose three evaluation methods.
\begin{itemize}
    \item Using $k$-means as a baseline, we can calculate the relative distances between the detected outliers with the centroids. If these values are small, then we conclude that our (LOF) method might not be a good method of anomaly detection. We call this ``Visualization Evaluation.'' 
    
    \item Since we represent our data in two ways with nodes and edges somehow exchanging, we can test if our methods are consistent by checking if detected suspicious users match with detected suspicious transactions. We call this ``Dual Evaluation."
    
    Specifically, with the user-as-node representation we can get the top $N$ user outliers and with the transaction-as-node representation we can get the top $M$ transaction outliers. In this paper, we choose $N = M = 100$. We then determine $X_N$ - the set of transactions corresponding to the top $N$ node outliers and $Y_M$ - the set of users corresponding to the top $M$ transaction outliers defined above. We define
    
\begin{equation*} 
    A_1 = \frac{|X_N \cap \text{ top $X_N$ transaction outliers}|}{|X_N|} 
\end{equation*}
and
\begin{equation*} 
    A_2 = \frac{|Y_M \cap \text{ top $Y_M$ user outliers}|}{|Y_M|}. 
\end{equation*}

Finally, we define the Dual Evaluation Metric $m_{DE}$ by
\begin{equation*} 
    m_{DE} = \frac{A_1 + A_2}{2}. 
\end{equation*}
Note that $m_{DE} \in [0, 1]$, and the bigger it is the more accurate our method is.
    
    \item Finally, there are around $30$ revealed thieves in the Bitcoin network. We can check if they belong to our detected suspicious user and transaction sets.
    
\end{itemize}

\section{Results}
We apply our network analysis techniques for both the network representations of the data. We implement these methods using the full dataset. 

\subsection{Power Degree $\&$ Densification Laws}
We first look at the densification power law for the two graph types. (See Figures $1-2$.) \\

\begin{figure}[h]
\begin{center}
\includegraphics[scale=0.39]{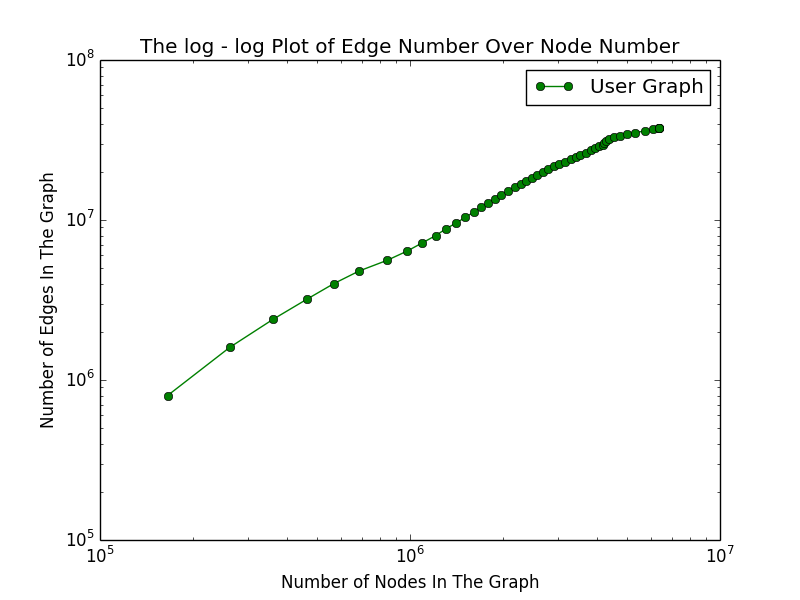}
\caption{The log-log Plot of Edge Number over Node Number - User Graph.}
\end{center}
\end{figure}

\begin{figure}[h]
\begin{center}
\includegraphics[scale=0.39]{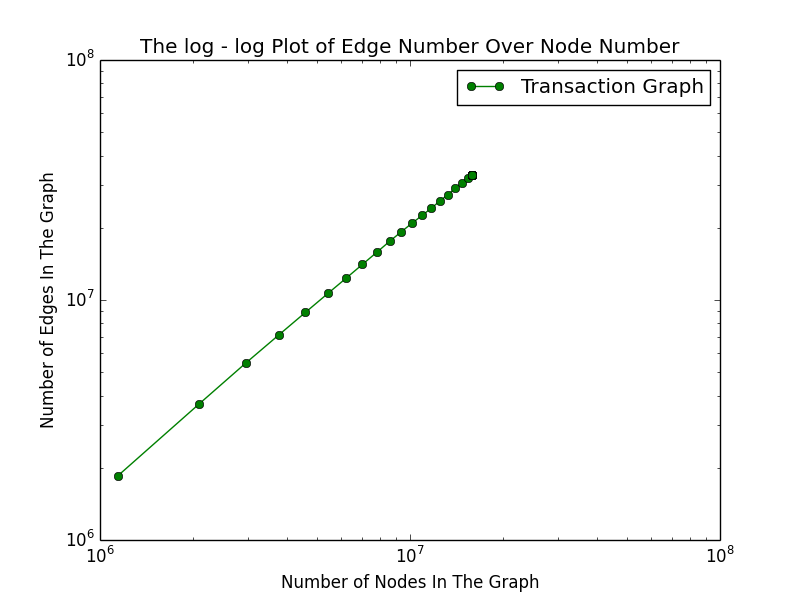}
\caption{The log-log Plot of Edge Number over Node Number - Transaction Graph.}
\end{center}
\end{figure}

The densification power law does not seem to be so well satisfied for the user-as-node graph, but perfectly satisfied for the transaction-as-node graph type.

A straight line for the transaction graph indicates that the number of edges (i.e. the number of transactions connecting with one another) is a reliable exponential function of the number of nodes (i.e. the number of transactions in the network) over time. This says, either there is nothing wrong with the network or people just try to hide the abnormal activities through seemingly legitimate transactions. If thieves are available, they will certainly put in every effort to make their transactions look licit in this easiest sense - the number of money flows among transactions is exponentially proportional to the number of transactions. Otherwise, they can easily be detected.

The corresponding plot for the user graph is different. In this plot, although the plot has the shape of a straight line, it bends in the middle and at the end of the line. This might indicate that some users enter the network and make an unusual number of transactions; in other words, thieves are well available. To explain for the first two plots, we observe that thefts in fact occur in the network; and while thieves can control for the legitimate appearance of the number of money flows among transactions, they cannot control for the number of transactions they make; when both of these are normal, they cannot win free money. 

Now, we investigate the power law degree distribution for various measures including in-degree, out-degree, balance, and average transaction size for the user-as-node graph type and in-degree, out-degree, total amount of transaction for the transaction-as-node graph. (See Figures $3-4$.)

\begin{figure}[h]
\begin{center}
\includegraphics[scale=0.39]{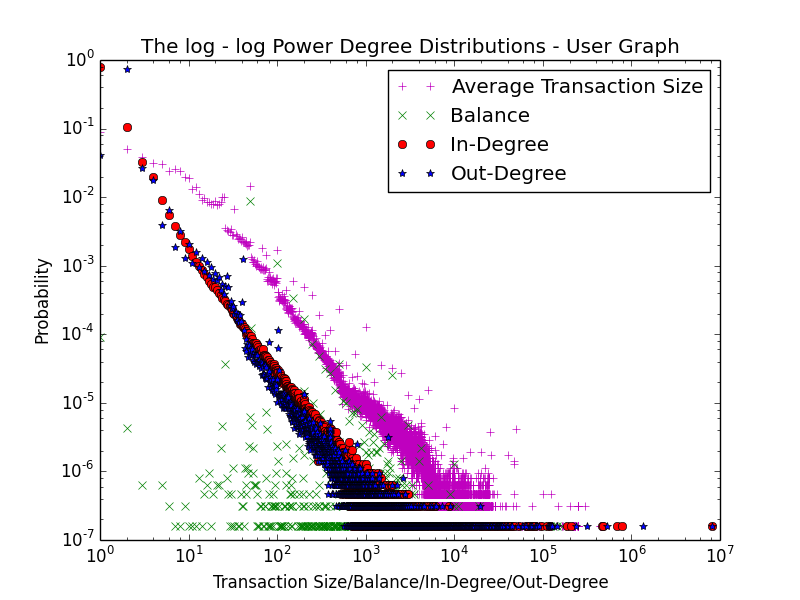}
\caption{The log-log Plot of Degree / Balance / Transaction Size Distributions - User Graph.}
\end{center}
\end{figure}

First, we observe the user graph. (Figure $3$.) The log-log plot of the in-degree distribution fits quite well the power degree law as there seems to be a straight line going through the points on the plot. So there would not be any problem with this distribution. Now we turn to the log-log plot of the out-degree distribution. The linear relation starts to appear broken as many points are unusually out of the core. As we have explained above, since thieves cannot control for the number of transactions they make (which correspond well to the degrees of nodes in the user graph) the power degree law should be violated for this distribution when there are thefts in the network. 

The log-log plots of the average transaction size and balance are even a more obvious sign of the existence of abnormal activities in the network. There appear to be many points in these plots which stay randomly and far away from the main core lines. These plots do not look fit for any usual kind of degree distribution and thus, the unusual points here correspond to suspicious users in the network. 

We now turn to the transaction graph. (Figure $4$.) 

\begin{figure}[h]
\begin{center}
\includegraphics[scale=0.39]{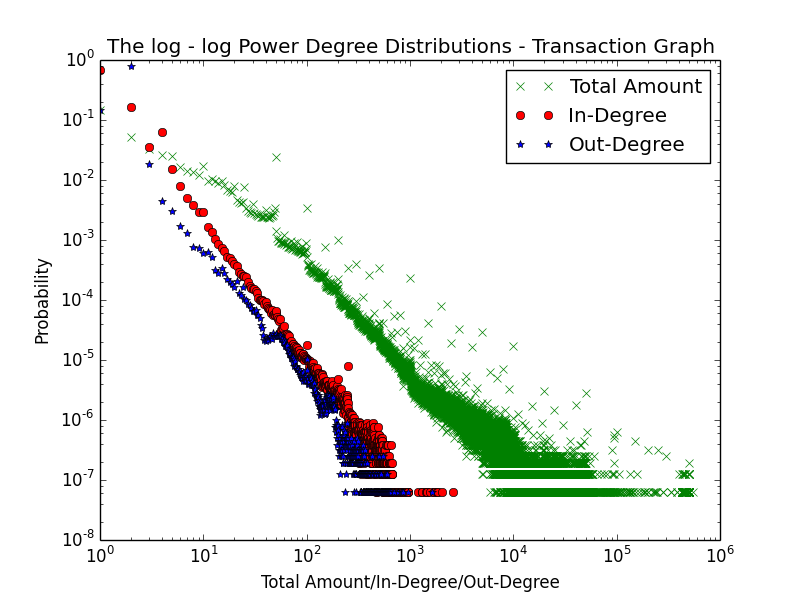}
\caption{The log-log Plot of Degree / Total Amount Distributions - Transaction Graph.}
\end{center}
\end{figure}

For this graph type the in-degree and out-degree appear quite alright, which makes sense with our straight line observation of the densification power law for transaction-as-node graph. However, the plot for total amount (for each transaction) behaves abnormally. There is a significant portion of points in the plot which stay far away from the core line; these points correspond to the suspicious transactions in the network.   

By the power degree $\&$ densification laws, we conclude that there are apparent anomalies in the network; in other words, there exist users who conduct illegal activities. Next, we will use $k$-means clustering method and local outlier factor method to figure out who those users might be.

\subsection{$k$-means Clustering}
Following Xiong et al. \cite{Xiong}, to find the optimal value of $k$ we calculate the entropy measure and choose the integer $k$ corresponding to the smallest entropy. Being mindful of computational time, we limit our selection of $k$ to be between $2$ and $10$. We run the clustering algorithm and find that for the user graph, at $k = 7$ we obtain the lowest entropy measure between clusters, which indicates that $k = 7$ is optimally selected for this graph type. (Figure $5$.) For the transaction graph, the optimal $k$ is $k = 8$. However, for the sake of the dual evaluation method, we will choose $k = 7$ for both graph types.

\begin{figure}[h]
\begin{center}
\includegraphics[scale=0.5]{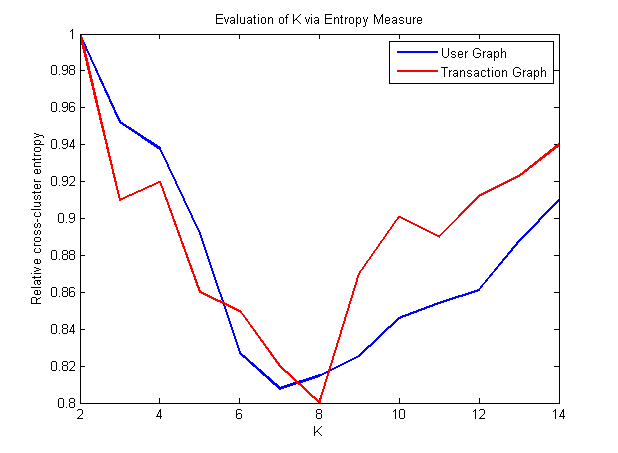}
\caption{Cross-cluster entropy versus $k$.}
\end{center}
\end{figure}

Now we plot these $k = 7$ clusters for each graph type. (Figures $6-7$.)

\begin{figure}[h]
\begin{center}
\includegraphics[scale=0.44]{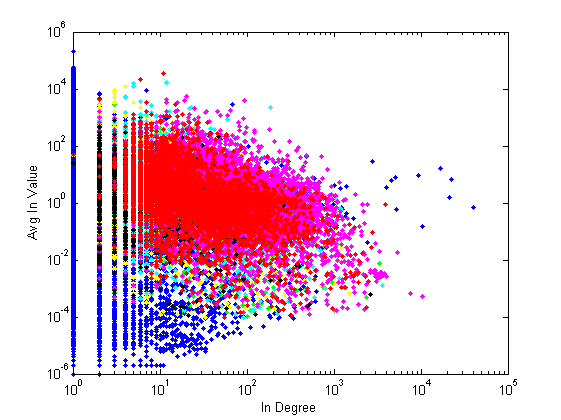}
\caption{$k$-means clustering of Average Inbound Value versus In - Degree. (User Graph)}
\end{center}
\end{figure}

\begin{figure}[h]
\begin{center}
\includegraphics[scale=0.44]{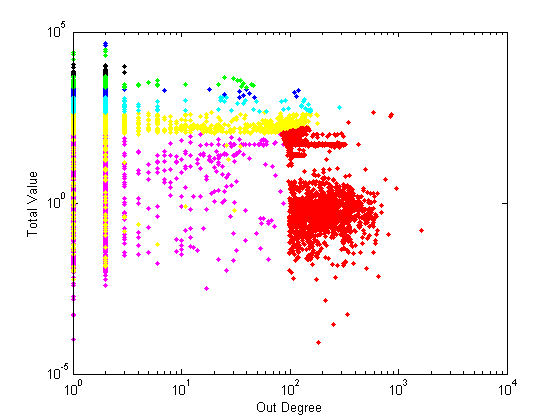}
\caption{$k$-means clustering of Total Value versus Out - Degree. (Transaction Graph)}
\end{center}
\end{figure}

Visualizing all points and their respective cluster assignments, we see that one cluster comprises the vast majority of the data points. The remaining six clusters are composed of users who are further away from this major cluster, but still comprise significantly large and significantly homogeneous groups in order to be clustered. Figures $6$ and $7$ are plots in two dimensions of the clusters found by our algorithm for two graph types. Figure $6$ is the plot for the user graph in two dimensions of in-degree and inbound value while Figure $7$ is the plot for the transaction graph with two dimensions being out-degree and total value. We can certainly plot over all pairs of features we use in $k$-means method, but for visualizing purpose we believe that these two plots are sufficient.

In Table $1$, we report the list of cluster centroids (for $k = 7$) for the user graph. (The corresponding table for the transaction graph is not quite illustrative so we will not present it here.) We find that all cluster centroids have relatively low cluster coefficients, implying that few users have high cluster coefficients, and the ones who do are not relatively similar to others. Other measures like in-degree, out-degree, mean in value, mean out value vary in values across the centroids. Hence, these measures not good indicators to tell whether some users are likely to be anomalies or not. 

\begin{table}
\centering
\begin{tabular}{@{}ccccc@{}} \hline
\multicolumn{1}{p{1.2cm}}{\centering In \\ Deg} & \multicolumn{1}{p{1.2cm}}{\centering Out \\ Deg} & \multicolumn{1}{p{1.2cm}}{\centering Mean \\ In Val} & \multicolumn{1}{p{1.2cm}}{\centering Mean\\  Out Val} & 
\multicolumn{1}{p{1.2cm}}{\centering Cluster \\ Coeff.} \\ \hline \hline
110.83 &	22.28 &	18.12 &	20.87 & 0.09 \\ 
10.48 &	2.36 &	28.86 &	47.77 & 0.07 \\ 
16.38 &	3.33 &	32.18 &	42.80 & 0.09 \\ 
54.52 &	10.44 &	26.72 &	27.20 & 0.07 \\ 
16.67 &	4.45 &	31.93 &	33.72 & 0.06 \\ 
5.66 &	5.88 &	184.70 &	184.45 & 0.03 \\ 
31.33 &	6.52 &	34.20 &	35.44 & 0.06 \\

\hline\end{tabular}
\caption{List of cluster centroids for $k = 7$. (User Graph)}
\end{table}

In the next part, we report the results obtained from the local outlier factor method. 

\subsection{Local Outlier Factor}
In Tables $2$ and $3$, we report a list of top nine outliers for the user and transaction graphs (i.e. anomalies) with $k = 7$. In each table, we rank the detected outliers (anomalies) in the reversed order of the LOF indices, starting with the highest value. The relative distance column gives the LOF indices relative to the maximum value of the LOF index for each graph type.

First, we consider the user graph. Looking at these outliers, one pattern we identify is that outliers all have a clustering coefficient of one. This could occur due to the scarcity of users who have such a high clustering coefficient, and so there is no centroid that has a higher clustering coefficient. The in-degree and out-degree measures vary in values, so they do not tell the story of anomalies. However, the mean-in-value and mean-out-value are quite interesting as it appears that the detected outliers have really similar values for these two measures. ($50$ for mean-in-value and $100$ for mean-out-value.) 

\begin{table}
\centering
\begin{tabular}{@{}cccccc@{}} \hline

\multicolumn{1}{p{1.2cm}}{\centering Relative \\ Distance} & \multicolumn{1}{p{0.8cm}}{\centering In \\ Deg} & \multicolumn{1}{p{0.8cm}}{\centering Out \\ Deg} & \multicolumn{1}{p{1.2cm}}{\centering Mean \\ In Val} & \multicolumn{1}{p{1.2cm}}{\centering Mean \\ Out Val} & 
\multicolumn{1}{p{1cm}}{\centering Cluster \\ Coeff.} \\
\hline \hline
1.00 &	3 &	3 &	47.6 &	81 &	1 \\ 
0.90 &	1 &	3 &	50 &	100 &	1 \\ 
0.20 &	4 &	4 &	47.5 &	85 &	1 \\ 
0.15 &	1 &	3 &	50 &	100 &	1 \\ 
0.14 &	3 &	5 &	50 &	100 &	1 \\ 
0.13 &	2 &	4 &	50 &	100 &	1 \\ 
0.13 &	1 &	3 &	50 &	100 &	1 \\ 
0.12 &	40 & 40 & 50 &	98 &	1 \\ 
0.12 &	1 &	3 &	50 &	100 &	1 \\

\hline\end{tabular}
\caption{List of top outliers for $k = 7$ (User Graph).}
\end{table}

Now, we consider the transaction graph. There are no obvious patterns for the detected outliers (i.e. suspicious transactions). We use three features for the transaction graph including in-degree, out-degree, and total amount of transaction; it appears that all these three measures vary in values. So similarly to the user graph, the in-degree and out-degree measures do not tell the story of anomalies. But differently from the user graph, in this case the total amount (value) measure is also not indicative of whether a node is abnormal.

\begin{table}
\centering
\begin{tabular}{@{}cccc@{}} \hline

\multicolumn{1}{p{1.2cm}}{\centering Relative \\ Distance} & \multicolumn{1}{p{0.8cm}}{\centering In Deg} & \multicolumn{1}{p{0.8cm}}{\centering Out Deg} & \multicolumn{1}{p{0.8cm}}{\centering Value} \\ \hline \hline
1.00 &	21 &	2 &	147.6 \\ 
0.98 &	20 &	3 &	102.1  \\ 
0.97 &	14 &	4 &	47.5  \\ 
0.77 &	12 &	2 &	50 \\ 
0.71 &	2 &	6 &	78.5 \\ 
0.65 &	9 &	2 &	89.3 \\ 
0.59 &	14 &	3 &	50  \\ 
0.57 &	40 & 40 & 89.5 \\ 
0.56 &	1 &	8 &	77  \\

\hline\end{tabular}
\caption{List of top outliers for $k = 7$ (Transaction Graph).}
\end{table}

We proceed by plotting the detected outliers using the LOF method together with $k$-means clusters for the two graph types. (Figures $8-9$.)

Figure $8$ is the plot for the user graph on two dimensions of the in-degree and average in-value. We observe that detected outliers (suspicious users) lie at the border surrounding the normal nodes. The nodes with extremely high value of the average in-value measure appear highly likely to be anomalies. The same conclusion is drawn for node with extremely low value of the in-degree measure. Some other detected outliers have values of these measures in between, though other measures should be really extreme instead.   

\begin{figure}[h]
\begin{center}
\includegraphics[scale=0.5]{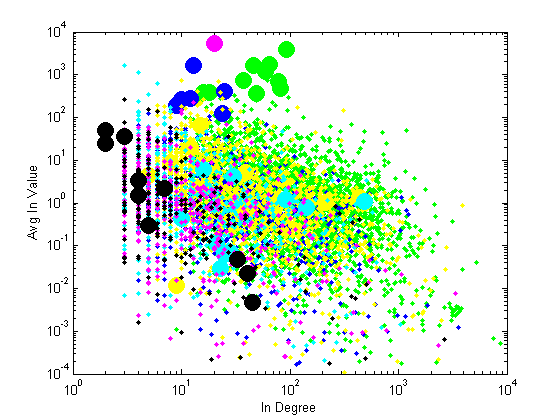}
\caption{$k$-means clustering of In - Degree versus Average In-Value. User outliers are represented by larger circles.}
\end{center}
\end{figure}

Figure $9$ is the plot for the transaction graph on two dimensions of the out-degree and total value. Similarly to the user graph, we observe that detected outliers (suspicious transactions) lie at the border of the plot. The nodes with extremely low and high value of the out-degree measure appear highly likely to be anomalies. Likewise, nodes with extremely high value of the total value (for each transaction) are likely to be detected as anomalies.

\begin{figure}[h]
\begin{center}
\includegraphics[scale=0.5]{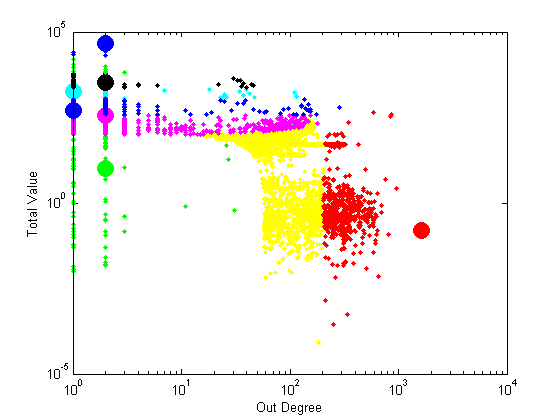}
\caption{$k$-means clustering of Total Value versus Out - Degree. Transaction outliers are represented by larger circles}
\end{center}
\end{figure}

Recall that in the result section of Power law Degree distribution $\&$ Densification power law, any measures related to money (balance, transaction size, total amount or value, etc.) behave abnormally with many extreme points staying out of the core shape of the plots. That observation is consistent with the detected outliers we get through the LOF method here in the sense that points with extreme values are likely to be anomalies. 

\section{Evaluation of Results}
We follow the three proposed evaluation methods to evaluate our findings. 

\begin{itemize}

    \item[\textbullet] Using $k$-means clustering method, we get $k$ clusters with corresponding $k$ centroids. For each graph type (for the LOF method), we calculate the average of the ratios of detected anomaly distances to corresponding centroids over max distances from those centroids to their assigned points for the top $100$ outliers.
    
    For the LOF method, we get $0.965$ for the user graph and $0.914$ for the transaction graph. These values are as large as we expect them to be since by observations in the Results section, detected anomalies appear to be extreme points.
    
    \item[\textbullet] To evaluate the agreement between the two graph types, we compute $A1 = 0.72$ and $A2 = 0.37$, resulting in $m_{DE} = 0.55$. The value of $m_{DE}$ is quite high, indicating that the LOF method is consistent.  

    \item[\textbullet] Using the LOF method, we detect one known theft via a Trojan keylogger that occurred in $2012$. The anomalous transaction was one in a series of transactions that obtained a total of $2,600$ Bitcoins from $22$ various addresses and funneled them to a single address.

\end{itemize}

\section{Future Studies}
The biggest challenge for problems with unlabelled data is testing. There is no obvious, universally agreed way to do testing for any kind of algorithms for such problems. In this project, besides using the publicly revealed thefts we propose two metrics with the hope to evaluate our methodology. Though with these evaluation methods we can somewhat evaluate the consistency of our (LOF) anomaly detection method, in no way we can evaluate its accuracy. 

In the future, we would like to go deeper into this issue. Specifically, we will find better evaluation methods. One direction is to generate similar data but with labels, and we test our methods on the newly generated data and finding results for the original data. 

Once better evaluation methods are proposed, we would also like to investigate better approaches to solving the anomaly detection problem. The LOF method is nice, but it is by no means the universally best one. In fact, literature has observed several modifications of this method as well as totally new ones in this area. We would like to look into the methods that can deal efficiently with network type data. 

Regarding the Bitcoin network, we can also improve on getting better features out of the data. One possible way is to segment transaction network into time slices. We can also detect smaller communities within the Bitcoin network and work explicitly within each community. 

\section{Conclusions}
In this paper, we investigate the Bitcoin network. We first represent the data with two focuses: users and transactions. We then use three main social network techniques to detect anomalies, which are potential anomalous users and transactions. While the agreement metrics are not high, we are able to detect one known case of theft, out of the 30 known cases we have. Since these techniques apply in other network settings, we in fact worked on a bigger problem of anomaly detection in networks.

\end{document}